\documentclass[onecollarge,natbib]{svjour2}
\bibpunct{[}{]}{;}{n}{}{,} 
\smartqed  
\usepackage{graphicx}
\graphicspath{{./figures/}}
\usepackage{amsmath}
\usepackage{braket}

\journalname{Few-body systems}
\begin{document}

\title{Combining few-body cluster structures with many-body mean-field methods}


\author{D. Hove \and E. Garrido \and A. S. Jensen \and P. Sarriguren \and H. O. U. Fynbo \and D. V. Fedorov \and N. T. Zinner}


\institute{D. Hove \and A. S. Jensen \and H. O. U. Fynbo \and D. V. Fedorov \and N. T. Zinner \at
Department of Physics and Astronomy, Aarhus University, 8000 Aarhus C, Denmark \\ \email{dennish@phys.au.dk}           
\and
E. Garrido  \and P. Sarriguren \at
Instituto de Estructura de la Materia, IEM-CSIC, Serrano 123, E-28006 Madrid, Spain
}

\date{Received: date / Accepted: date}

\maketitle

\begin{abstract}
Nuclear cluster physics implicitly assumes a distinction between groups of degrees-of-freedom, that is the (frozen) intrinsic and (explicitly treated) relative cluster motion. We formulate a realistic and practical method to describe the coupled motion of these two sets of degrees-of-freedom. We derive a coupled set of differential equations for the system using the phenomenologically adjusted effective in-medium Skyrme type of nucleon-nucleon interaction. We select a two-nucleon plus core system where the mean-field approximation corresponding to the Skyrme interaction is used for the core. A hyperspherical adiabatic expansion of the Faddeev equations is used for the relative cluster motion. We shall specifically compare both the structure and the decay mechanism found from the traditional three-body calculations with the result using the new boundary condition provided by the full microscopic structure at small distance. The extended Hilbert space guaranties an improved wave function compared to both mean-field and three-body solutions. We shall investigate the structures and decay mechanism of $^{22}$C ($^{20}$C+n+n). In conclusion, we have developed a method combining nuclear few- and many-body techniques without losing the descriptive power of each approximation at medium-to-large distances and small distances respectively. The coupled set of equations are solved self-consistently, and both structure and dynamic evolution are studied.

\end{abstract}

\section{General approach \label{sec:approach}}

\begin{figure}
\centering
\begin{minipage}{.4\textwidth}
	\includegraphics[width=1\linewidth]{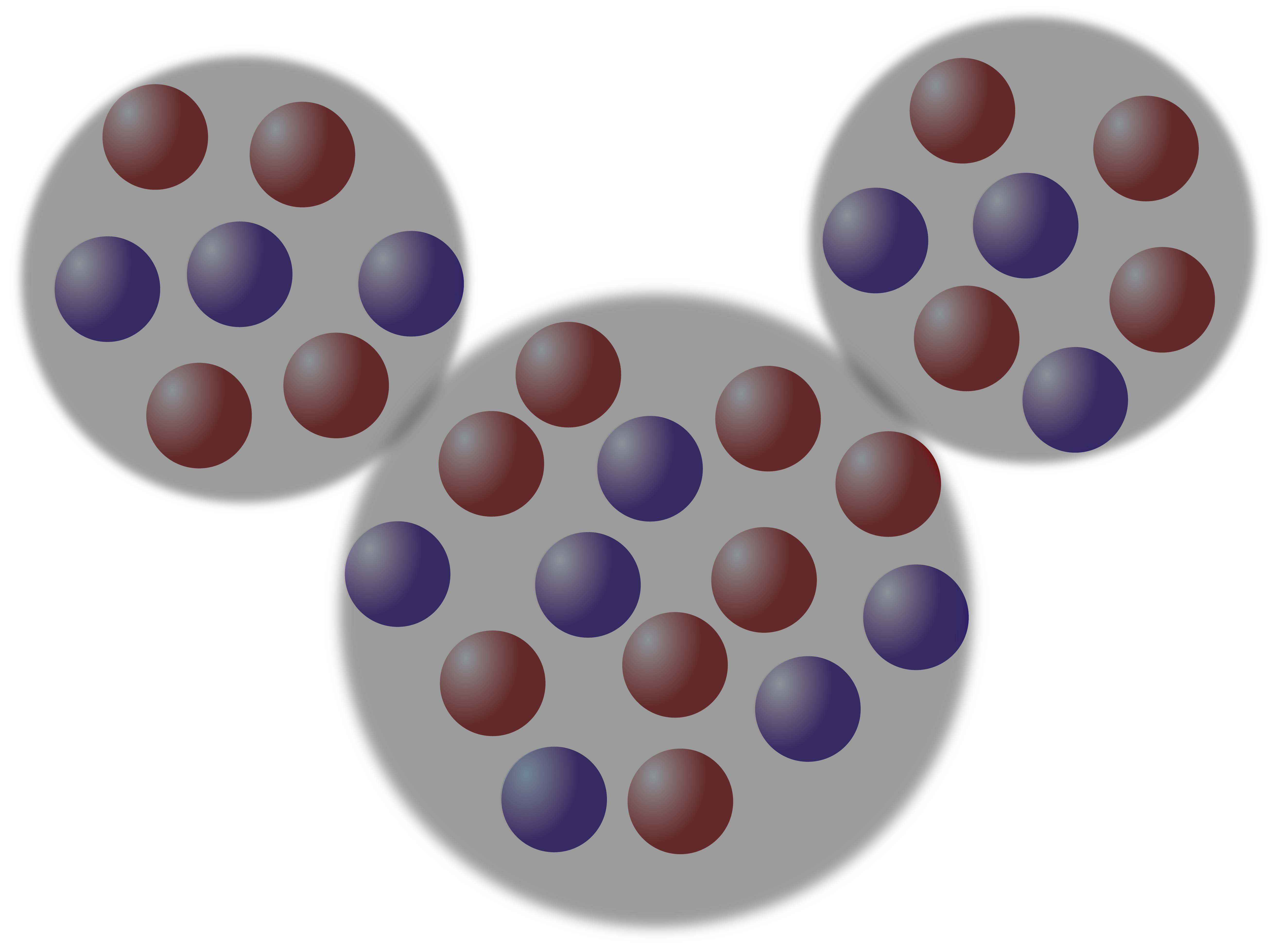}
\end{minipage}\hfill
\begin{minipage}{0.55\textwidth}
	\caption{A many-body system with $A$ particles interaction through some interaction. Because of the arrangement a three-body cluster structure can be assumed (superimposed in grey). \label{fig:manybody}}
\end{minipage}
\end{figure}

Whenever one attempts to describe a physical system the first choice must always be a choice of the relevant degrees of freedom. In a system consisting of $A$ particles with some interaction one could apply some form of many-body description, depending on the system in question. This is often very useful, but it can make it difficult to treat long-range effects or breakup reactions. However, if we have some physical insight or some experimental evidence indicating a specific structure, one can often reasonably assume an appropriate clusterization \cite{jen04}. This is illustrated  schematically in Fig.~\ref{fig:manybody}.

With an assumed few-body cluster structure it is much easier to treat exotic systems, such as borromean or halo systems \cite{riis13,hov14}. It is also easier to describe long-range effects and dynamic breakups, simply because all particles are not treated equally \cite{hov16}. The main problem with clusterized models is the difficulty in determining the effective cluster interactions. As it is very difficult to deduce the effective interaction from fundamental interactions, one often ends up having to create them phenomenologically, adjusted to some physical observable of interest. This severely limits the scope of cluster models in situations where experimental information is lacking. Another more intuitive problem follows from surrendering that many degrees of freedom. This suggests a reduction in the level of detail possible.

Our intent is to combine the two conceptually different pictures, and produce a model with the flexibility of traditional cluster models, but with the level of detail of a many-body description. More importantly, we aim to establish a method for computing the effective interactions in the cluster model from more general interaction parameters. The fundamental idea is to view the system as a three-body system, superimposed on a many-body system, as illustrated in Fig.~\ref{fig:manybody}.

The starting point is a many-body Hamiltonian corresponding for instance to the many-body system illustrated in Fig.~\ref{fig:manybody}. The total system consists of $A$ particles, while the number of particles in the $i$'th cluster is $A_i$. This Hamiltonian can be expressed as 
\begin{align}
H &= \sum_{i=1}^A T_i + \frac{1}{2} \sum_{i,j}^{A} V_{ij} + \frac{1}{6} \sum_{i,j,k}^{A} V_{ijk},
\label{eq:ham_many}
\end{align}
where $T_i$ is the kinetic part of the $i$'th particle, while $V_{ij}$ and $V_{ijk}$ are two- and three-body interactions respectively. Higher order contributions are assumed to be negligible. The various sums can be separated into terms relating to each of the three clusters as well as the interaction between particles in different clusters. The resulting Hamiltonian is 
\begin{align}
H &= H_1 + H_2 + H_3 
+ \sum_{i}^{A_1} \sum_{j}^{A_2} \frac{V_{ij}}{2}
+ \sum_{i}^{A_1} \sum_{j}^{A_3} \frac{V_{ij}}{2}
+ \sum_{i}^{A_2} \sum_{j}^{A_3} \frac{V_{ij}}{2}
+ \sum_{i,j}^{A_1} \sum_{k}^{A_2} \frac{V_{ijk}}{6}
+ \sum_{i,j}^{A_1} \sum_{k}^{A_3} \frac{V_{ijk}}{6}
\notag \\
&+ \sum_{i,j}^{A_2} \sum_{k}^{A_1} \frac{V_{ijk}}{6}
+ \sum_{i,j}^{A_2} \sum_{k}^{A_3} \frac{V_{ijk}}{6}
+ \sum_{i,j}^{A_3} \sum_{k}^{A_1} \frac{V_{ijk}}{6}
+ \sum_{i,j}^{A_3} \sum_{k}^{A_2} \frac{V_{ijk}}{6}
+ \sum_{i}^{A_1} \sum_{j}^{A_2} \sum_{k}^{A_3} \frac{V_{ijk}}{6}.
\label{eq:ham_sep}
\end{align}
To begin with a simple system is chosen, where two of the clusters are assumed to consist of one particle each (valence particles), and that these particles are identical. To avoid higher order terms the three-body interaction is parameterized as a density dependent two-body interaction, which is absorbed into the various two-body interaction terms. The Hamiltonian then simplifies to 
\begin{align}
H = \left( H_{core} \right)
+ \left( T_{v_1} + T_{v_2} + V_{v_1 v_2} + V_{3b} \right)
+ \left( \sum_{i}^{A_1} V_{iv_1} + \sum_{i}^{A_1} V_{iv_2} \right)
= H_{core} + H_{3b} + H_{mix}, 
\label{eq:Ham_simp}
\end{align}
where $H_{core}$ is the Hamiltonian for the remaining cluster (the core), $T_{v_{1/2}}$ is the kinetic term for the valence particles, $V_{v_1 v_2}$ is the two-body interaction between the valence particles, $V_{iv_{1/2}}$ is the two-body interaction between a valence particle and the core particles, and $V_{3b}$ is the three-body interaction between the two valence particles and the core particles. In this way the Hamiltonian separates into a part relating to core degrees-of-freedom, $H_{core}$, a part (second parenthesis) relating to three-body degrees of freedom, $H_{3b}$, and a part (third parenthesis) mixing both degrees-of-freedom, $H_{mix}$.

Having established a Hamiltonian all we need to do is to pick a wave function, pick an interaction, and do a variation. Given that the fundamental picture is a three-body structure superimposed on a many-body system a natural choice of wave function is an antisymmetric product of a core wave function, $\psi_c$, and a three-body wave function, $\psi_{3b}$,
\begin{align}
\Psi = \mathcal{A}\left(\psi_c(\mathbf{r}_1, \dots , \mathbf{r}_{A_1}) \psi_{3b}(\mathbf{r}_{v_1},\mathbf{r}_{v_2}) \right),
\label{eq:wavefunc}
\end{align}
where $\mathcal{A}$ indicates the anti-symmetrisation of all particles, while $\mathbf{r}_i$ and $\mathbf{r}_{v_i}$ are the spin and space coordinates of the $i$'th core or valence particle. Without loss of generality traditional Jacobi coordinates can be used for the three-body wave function.

In general, varying the energy with respect to both the core and the three-body wave function leads to a set of coupled differential equations 
\begin{align}
E_c \psi_c 
&= H_{core} \psi_c + \bra{\psi_{3b}} H_{mix} \psi_c \ket{\psi_{3b}}
+ \Braket{\psi_{3b} |  \psi_c^{\ast} \frac{\delta H}{\delta \psi_c} \psi_c | \psi_{3b}},
\\
E_{3b} \psi_{3b}
&= H_{3b} \psi_{3b} + \bra{\psi_c} H_{mix} \psi_{3b} \ket{\psi_c} 
+ \Braket{ \psi_c | \psi_{3b}^{\ast} \frac{\delta H}{\delta \psi_{3b}} \psi_{3b} | \psi_c}.
\end{align}
The final term in each equation appears because the Hamiltonian does not separate completely due to the density dependency of the two-body interaction.

All this has been very general. No specific system has been chosen and no specific interaction has been specified. If the test system is a nuclear system, the most natural core wave function would be a Slater determinant of single-particle states. Our corresponding interaction of choice is a simple density dependent Skyrme interaction. The coupled equation then takes the form
\begin{align}
\epsilon_i \psi_i
&= \left( T + V_C(\psi_{3b}) + V_{SO}(\psi_{3b}) + \vec{\nabla} \cdot m_{eff}(\psi_{3b}) \vec{\nabla} \right) \psi_i, 
\label{eq:core}\\
E_{3b} \psi_{3b}
&= \left( T + V_C(\psi_i) + V_{SO}(\psi_i) + \vec{\nabla} \cdot m_{eff}(\psi_i) \vec{\nabla} \right) \psi_{3b}. \label{eq:3b}
\end{align}
The interactions in Eqs.~(\ref{eq:core}) and (\ref{eq:3b}) both consist of central, $V_C$, and spin-orbit, $V_{SO}$, potentials as well as effective mass terms. 

Within the world of mean-field Hartree-Fock calculations with Skyrme forces these interactions are completely standard. Our many-body method is therefore identical to the traditional method championed by Vautherin and others \cite{vau72}, only the specific parameters in the interaction have changed due to the effect of the external valence nucleons. Likewise, the interaction between the core and valence nucleons in both Eqs.~(\ref{eq:core}) and (\ref{eq:3b}) is also completely determined by the Skyrme parameters. However, as Skyrme interactions are in-medium interactions, the interaction between valence nucleons is the free parametrized interaction from Ref.~\cite{gar04}. Within the world of three-body cluster methods effective masses are rather uncommon, but they can be handled. Our three-body method is the hyperspheric adiabatic expansion of the Faddeev equations in coordinate space \cite{nie01}.

Generally, traditional Hartree-Fock calculations yield very good results, especially with regards to ground state binding energy, but they tend to struggle with more exotic systems and in regions of the nuclear chart where rapid structural changes occur \cite{hov14b}, where the individual behaviour of the final nucleons becomes more important. This is better accounted for when combining the the mean-field methods with the clusterized three-body structure. The main point of the described method is that the effective cluster interaction is being derived externally. Having chosen a set of Skyrme parameters the only remaining freedom is the three-body interaction. Of course it is unrealistic to expect a single set of Skyrme parameters to be useful for our purposes for the entire nuclear chart. However, the hope is not only that a single set of Skyrme parameters might yield a more realistic result for a given three-body system, but also that a single parameter set could be used in an extended region of the nuclear chart. 

Even though there is an added level of complexity with this method, any observable that could be calculated with either traditional mean-field or traditional three-body methods can be calculated now. This provides a very useful frame for comparing the results. Not only can the calculations be compared against traditional Hartree-Fock calculations, they can also be compared against traditional three-body calculations. The comparison will here be a structural comparison against a traditional three-body calculation.

\section{Short-range structure and applications \label{sec:struc}}

Given that we change the two-body interaction, we necessarily change the short-range boundary condition. At the very least this will manifest as short-ranged structural changes. However, seeing as the short-range structure dictates the long-range behaviour along with the breakup mechanisms, this change in structure will have profound influence on the calculation of many physical observables. 

\begin{figure}
\centering
\begin{minipage}{.49\textwidth}
	\includegraphics[width=1\linewidth]{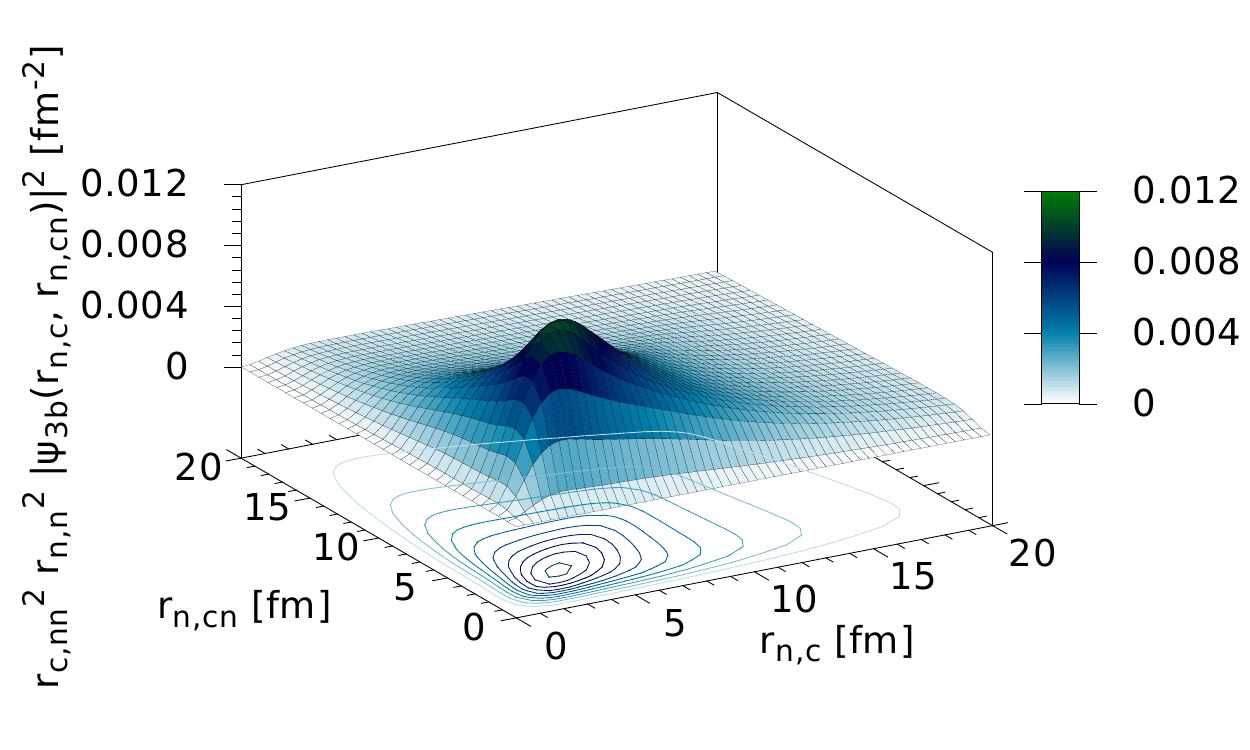}
\end{minipage}\hfill
\begin{minipage}{.49\textwidth}
	\includegraphics[width=1\linewidth]{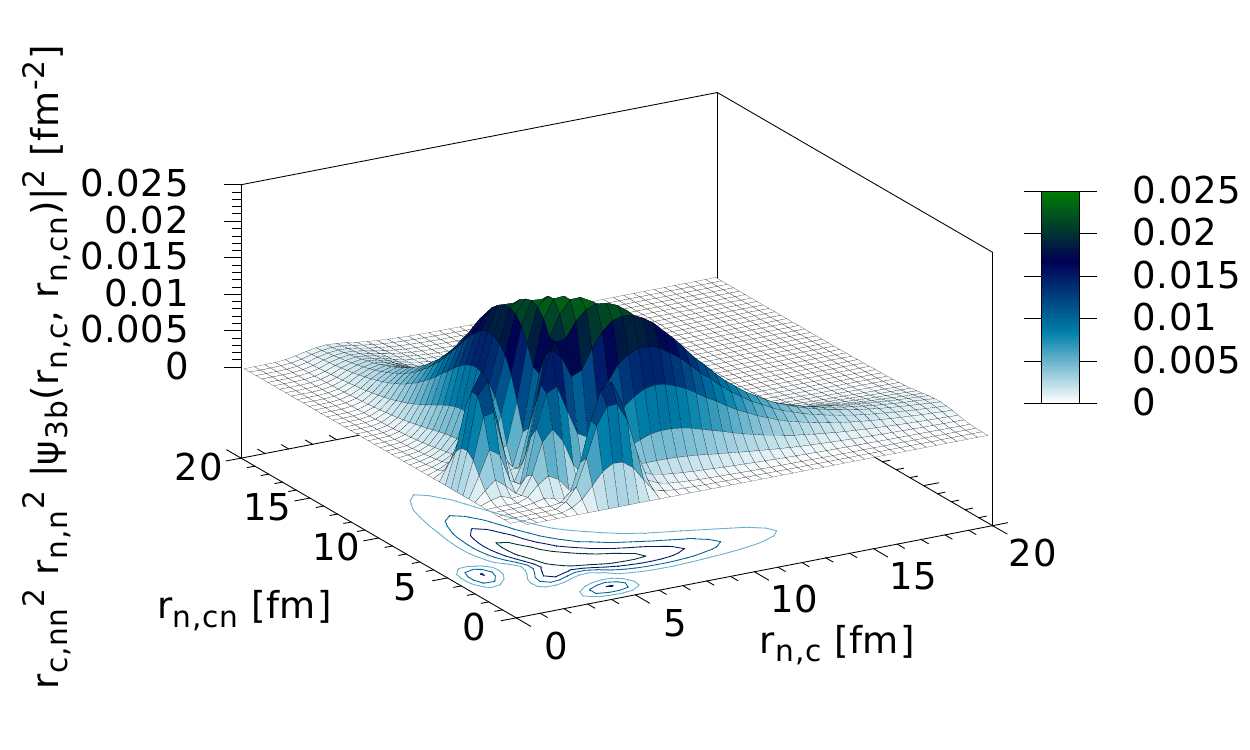}
\end{minipage}\hfill
\caption{(Left) The square of the three-body wave function $\psi_{3b}$ for $^{20}C + n + n$  calculated using traditional three-body methods. See Ref.~\cite{nie01} for more details. The coordinates are relative coordinates between core and neutron, and between neutron and core-neutron system. (Right) The same calculated using our method, where the two-body interaction is determined by the underlying many-body structure. \label{fig:3bwave}}
\end{figure}

Our test system is $^{22}\text{C}$ viewed as a $^{20}\text{C}$ and two neutrons. This system is right at the edge of the neutron dripline, where the size of the pairing \cite{hov13} makes $^{21}$C unbound, but $^{22}$C bound. The ability to treat the last neutrons individually is therefore critical. The square of the three-body wave function for a traditional three-body calculation of $^{22}C$ viewed as $^{20}C$ and two neutrons is seen in the left pane of Fig.~\ref{fig:3bwave}. The same only based on our method is seen in right pane. The wave function is shown as a function of relative coordinates between the core and one valence neutron, $r_{n,c}$, and between the remaining valence neutron and the core-neutron system, $r_{n,cn}$.

%

From Fig.~\ref{fig:3bwave} a clear structural difference is seen. In the traditional calculation the system forms an equidistant configuration with both neutrons being roughly the same distance from the core. In our calculation the one neutron is situated close to the core and the other further away.

As mentioned this short range structural difference will also manifest at long range, where the predicted breakup mechanisms would be very different. The traditional calculation would lead to a combination of sequential and direct decay, while our calculation would lead to a very sequential decay. This will be elaborated further in coming publications.

To determine whether this new calculation in general is more realistic it would be necessary to compare to physical observables. For this specific system the most obvious candidates are the momentum distribution following one- or two-neutron removal \cite{kob12} or the invariant mass spectrum of the system. These are some of our objectives in the near future.

\end{document}